\documentclass[preprint,showpacs,preprintnumbers,amsmath,amssymb]{revtex4}
\usepackage{graphicx}
\usepackage{dcolumn}
\usepackage{bm}
\usepackage{color}

\begin{document}

\title{How the site degree influences quantum probability on inhomogeneous substrates}


\author{A. M. C. Souza$^{1}$, R. F. S. Andrade$^{2}$, N. A. M. Ara\'{u}jo$^{3,4}$, and H. J. Herrmann$^{5,6}$}

\affiliation{
$^{1}$Departamento de F\'{i}sica, Universidade Federal de Sergipe 49100-000, Sao Cristovao, Brazil.\\
$^{2}$Instituto de F\'{i}sica, Universidade Federal da Bahia, 40210-210, Salvador, Brazil.\\
$^{3}$Departamento de F\'{\i}sica, Faculdade de Ci\^{e}ncias, Universidade de Lisboa, P-1749-016 Lisboa, Portugal.\\
$^{4}$Centro de F\'{i}sica Te\'{o}rica e Computacional, Universidade de Lisboa, P-1749-003 Lisboa, Portugal.\\
$^{5}$Computational Physics, IfB, ETH-H\"{o}nggerberg, Schafmattstr. 6, 8093, Z\"{u}rich, Switzerland.\\
$^{6}$Departamento de F\'{i}sica, Universidade Federal do Cear\'{a}, Campus do Pici, 60455-760, Fortaleza, Brazil.}

\date{\today}

\begin{abstract}
We investigate the effect of the node degree and energy $E$ on the electronic wave function for regular and irregular structures, namely, regular lattices, disordered percolation clusters, and complex networks. We evaluate the dependence of the quantum probability for each site on its degree. For bi-regular structures, we prove analytically that the probability $P_k(E)$ of finding the particle on any site with $k$ neighbors is independent of $E$. For more general structures, the dependency of $P_k(E)$ on $E$ is discussed by taking into account exact results on a one-dimensional semi-regular chain:  $P_k(E)$ is large for small values of $E$ when $k$ is also small, and its maximum values shift towards large values of $|E|$ with increasing $k$. Numerical evaluations of $P_k(E)$ for two different types of percolation clusters and the Apollonian network suggest that this feature might be generally valid.
\end{abstract}

\pacs {64.60.ah, 64.60.aq, 71.23.A, 72.15.Rn}

\maketitle

\section{Introduction}

The electronic conduction is one of the most important properties of a solid. It depends essentially on the localized/delocalized character of the electronic wave function, which is related to the intrinsic properties of the atoms in the material and its crystalline structure. As well known, the presence of disorder changes the extended character of the electronic states in periodic lattices, as established by Bloch's theorem \cite{ashcroft1976}. The introduction of disorder (substitutional, vacancies, etc.) is the main mechanism controlling the Anderson transition \cite{anderson1958}. Since disorder may emerge in many ways, different disorder types introduced on regular lattices produce different kinds of localized states \cite{evers2008}.

In the case of substitutional disorder, a large number of results obtained on different systems indicates that the wave function has a strong tendency to be localized on the sites occupied by defects having a number of connections that significantly differs from the lattice average coordination \cite{farkas2001,jahnke2008,giraud2009}. The way this general property is manifested still depends largely on the detailed substitutions, as well as on the energy of the eigenstates. Thus, many issues remain open in understanding how the lattice structure \cite{schreiber1996}, disorder \cite{moura1998}, and eigenstate energy favors the wave function localization on particular sets of sites, and on the possibility of controlling the wave function localization \cite{islam2009,darazs2013,chandrashekar2014}.

Usually, the investigation of the effect of disorder on localization of a given system is targeted at the construction of the phase diagram in terms of the energy and a disorder control parameter, where the transition from the extended to the localized states can be clearly identified. Several global properties characterizing extended and localized states can also be obtained as function of the quoted parameters.

In this work, we focus our investigation on the relation between the degree of a given site in a inhomogeneous structure and the amplitude of the wave function in its neighborhood. We restrict our study to purely geometrical features, and consider clusters on the square lattice obtained by usual bond percolation \cite{broadbent1957,stauffer1994}, Gaussian percolation \cite{araujo2010}, as well as on the deterministic Apollonian network \cite{andrade2005}, and some linear, periodic, inhomogeneous chains \cite{macedo1995}. We carry out a numerical investigation of the local properties of the wave function of a system described by a nearest neighbor tight-binding Hamiltonian.

For both types of percolation on the square lattice of side $L$, subject to periodic boundary conditions, with $N=L\times L$ sites, $N_b$ connections, $p=N_b/2N$ indicates the probability of a having a bond between nearest-neighbors sites. The models differ by the algorithm used to select bonds and, for the same value of $p$, the number of sites with $k$ neighbors, with  $k\in[0,4]$, may be different. We consider values of $p$ close to and above the percolation threshold $p_c$. We consider the percolation cluster which is a fractal of fractal dimension 91/48. According to current investigations, all wave functions at $p\gtrsim p_c$ are localized. The Apollonian network is characterized by a scale-free distribution of node degree and, as a consequence, the degree (number of connections) of a site varies in a wide interval. For this network, previous studies have indicated the presence of extended and localized states \cite{souza2008}. Finally, inhomogeneous chains are used to obtain exact results that help the discussion of more complex structures.

The results for the local probability distribution as a function of the degree hints at further steps towards controlling wave localization at a given site on the percolation cluster.

The rest of this work is organized as follows: In Sec. II, we present the electronic and two percolation models used to model a disordered system, and introduce two measures to characterize the electronic localization. Section III discusses exact results for semi-regular  lattices, which provide a useful comparison for the analysis that is presented for complex networks in Sec. IV and disordered systems in Sec. V. Finally, Sec. IV closes the paper with some final remarks on how the obtained results can be extended to more complex geometries.

\section{The model}

The simplest model for electric conduction is based on a one-particle tight-binding Hamiltonian. If we consider an ordered system on a periodic Bravais lattice, the electron interacts with the atom at any lattice site $\vec{r}$ with on-site energy $\epsilon_{\vec{r}}$, but it only jumps from $\vec{r}$ to site $\vec{r'}$ with a hopping probability $V_{\vec{r},\vec{r'}}=V_{\vec{r'},\vec{r}}$ if the two sites satisfy some conditions. Usually, it is assumed that $V_{\vec{r},\vec{r'}}=0$ unless $\vec{r}$ and $\vec{r'}$ are next neighbors and, if this is satisfied, $V_{\vec{r},\vec{r'}}$ is a constant independent of the pair of interacting sites.  This is justified by the fact that the hopping term results from the overlap integral of two one-particle wave-functions localized on neighboring sites mediated by an interaction potential. Thus, the system Hamiltonian is written as

\begin{equation} \label{eq1}
\mathbf{H}=\sum_{\vec{r}} \epsilon_{\vec{r}} |\vec{r}\rangle \langle \vec{r}
| + \sum_{(\vec{r},\vec{r'})}  V_{\vec{r},\vec{r'}} |\vec{r}\rangle \langle
\vec{r'}|.
\end{equation}

The eigenstates of $\mathbf{H}$ corresponding to eigenvalue $E(\vec{\kappa})$ are denoted by $|\psi_E \rangle$, so that the solutions of the Schr\"{o}dinger equation are the wave function $\psi_{E}(\vec{r})= \langle \vec{r}|\psi_E \rangle $, where $\vec{\kappa}$ denotes the wave vector so that $E=E(\kappa)$.

The Hamiltonian (\ref{eq1}) can also be extended to describe disordered systems on percolating clusters or on inhomogeneous substrates like the Apollonian network. This network is obtained through the recursive application of a geometrical procedure, which leads to a series of network generations identified by a label $g$. Thus, the investigation of the tight-binding system requires the use of a sequence of Hamiltonian operators $\mathbf{H_g}$ that account for all interactions between the available sites that have been introduced until generation $g$. Details of this kind of investigation can be found in Refs. \cite{souza2008,lyra2009}, where an investigation of the properties of the eigenstates for successive generations of the Apollonian network has been carried out. The site labeling used herein has been introduced in Ref. \cite{souza2008}.

In all cases we consider here, the local energy parameter $\epsilon_{\vec{r}}$ is assumed to be constant and, without loss of generality, is set to zero, what allows us to concentrate on the effect of topological disorder. For the percolation clusters, we consider $V_{\vec{r},\vec{r'}}=V_0$ or $V_{\vec{r},\vec{r'}}=0$, according to whether a bond between the sites $\vec{r}$ and $\vec{r'}$ is present or not. For inhomogeneous substrates, $V_{\vec{r},\vec{r'}}=V_0=1$ or $V_{\vec{r},\vec{r'}}=0$ according to whether the corresponding sites are connected or not.

Like in the classical percolation transition, the probability $p$ is also the control parameter for the the localized-extended transition, which some times is referred to as quantum percolation   \cite{harris1984,odagaki1984}. A quantum percolation threshold $p_q$ can be defined as the smallest value of $p$ for which there exists an eigenvalue $E$ of the Schr\"{o}dinger equation such that $|\psi_E \rangle$ is an extended eigenstate in the sense that it is not possible to find any finite region such that the sum of $|\psi_{E}(\vec{r})|^2$ over all sites outside this region is smaller than any arbitrarily chosen  positive number. The critical values $p_q$ for the quantum problems are usually larger than the corresponding percolation transition values $p_c$. For instance, for the square lattice, the bond percolation transition occurs at $p_c=0.5$. However, great controversy persists about the precise value of the quantum percolation threshold  \cite{chang1995,kantelhardt1998,gong2009,albuquerque2012}.

\newpage

In the usual random percolation problem, an empty connection is randomly chosen to be occupied in a sequential order. To obtain percolation clusters according to the Gaussian model \cite{araujo2010}, one starts with a regular lattice without bonds. The original bonds of the corresponding Bravais lattice are selected uniformly at random and added to the lattice with probability $q$, given by

\begin{equation}\label{eq2}
q=\mbox{min}\left\{1, \exp\left[-\alpha \left(\frac{s-\bar{s}}{\bar{s}}\right)^2\right]\right\}, \ \
\end{equation}

\noindent where $s$ is the size of the cluster of connected sites that will be formed if the bond  is added to the system and $\bar{s}$ is the average cluster size. Differently from the usual percolation clusters, the clusters that emerge at $p_c$ with the Gaussian rule are compact with a fractal perimeter.

To characterize the localization of the time independent wave function, we analyze the participation ratio ($\xi$) associated to the eigenvectors of the eigenvalues $E$, which has been successfully used to characterize the localization of the wave function for a large variety of systems. It is generally defined as  \cite{souza2008}

\begin{equation}\label{eq3}
  \xi(E) = \frac{1}{\sum_{\vec{r}} |\psi_{E}(\vec{r})|^{4}},
\end{equation}

\noindent where $\psi_{E}(\vec{r})$ is the wave function amplitude on site $\vec{r}$. A localized eigenstate is characterized by $\xi(E)/N\rightarrow 0$ in the limit $N \rightarrow \infty$, but this ratio converges to a finite value if the state is extended. To characterize the $k$-dependent probability of finding the particle at a site of degree $k$, we evaluate

\begin{equation}\label{eq3a}
  P_{k}(E) = \sum_{\vec{r}} |\psi_{E}(\vec{r})|^{2}\delta_{k(\vec{r}),k},
\end{equation}

\noindent with $\sum_{k} P_{k}(E)=1$.

Our investigation is based on the systematic evaluation of the relation between $\psi_{E}(\vec{r})$ and the site degree $k(\vec{r})$, from which the distribution $P_k(E)$ can be evaluated. We remark that, to the best of our knowledge, no similar results to those we report in this work have been reported before. As we will show in the next sections, the definition in Eq. (\ref{eq3a}) provides useful insights on the distribution probability of finding the particle on sites with different values of $k$ as a function of the energy.

In the case of percolation clusters, we restrict our analysis to those clusters evaluated at $p\gtrsim p_c$ as well for  the usual random occupation algorithm as for the Gaussian model of discontinuous percolation. In the case of the Apollonian network, we consider the situation in which all connections defined by the construction procedure correspond to a single value of the hopping integral.

\section{Semi-regular networks}

Before presenting the results of our simulations, let us briefly discuss some aspects of wave function localization in some simple structures, which helps understanding the behavior observed for more complex geometrical arrangements. We start with the general concept of graphs as a set $G=\{V,E\}$, where $V$ and $E$ stand, respectively, for the set of vertices (or nodes, or sites) and for the set of edges (or connections) between vertices. For the sake of clarity, networks where each site has the same number of neighbors (degree) are called \emph{lattices}, and are viewed as subset of \emph{regular} networks. Periodic networks with two sets of sites (say $S_1$ and $S_2$) are called \emph{semi-regular} networks when the degree of all sites in $S_1$ is $k_1$ and the degree of all sites in $S_2$ is $k_2$. On the other hand, networks are \emph{bipartite} when we can split the sites into two disjoint sets, in such a way that sites in set $S_1$ ($S_2$) are only directly connected to sites in set $S_2$ ($S_1$)  \cite{albert2002}. A trivial example of a bipartite lattice is the square lattice that can be split into two sublattices with the above property. It is easy to observe that the bipartite concept can be easily extended to include a larger number of partitions (e.g., the face centered cubic lattice is quadripartite). If we consider all semi-regular networks, it is possible to cast them into two different classes according to the following criterion: semi-regular networks that are also bipartite are called \emph{bi-regular} networks, which we identify as class $A$. Semi-regular but non-bipartite networks belong to class $B$, and are simply called semi-regular. Class $A$ satisfies the following condition: the product of the site degree of each set by the number of sites in this set is the same for the two sets. On the other hand, semi-regular networks in class $B$ do not satisfy this condition. For instance, the square lattice belongs to class $A$.

An important analytical result can be derived for any bi-regular structure of class $A$ (see Appendix): the probability of finding an electron (square modulus of the wave function) on a particular site type is independent of the wave function moment (or energy). On the other hand, it is easy to find counter examples of semi-regular structures of class $B$ showing that this result does not hold. Hence, the exact results for this very general topological classification can be used to explain specific wave-function properties for tight-binding models built on both regular Euclidian lattices and complex networks.

\begin{figure}[h!]
\centering
\includegraphics[width=10cm,angle=0]{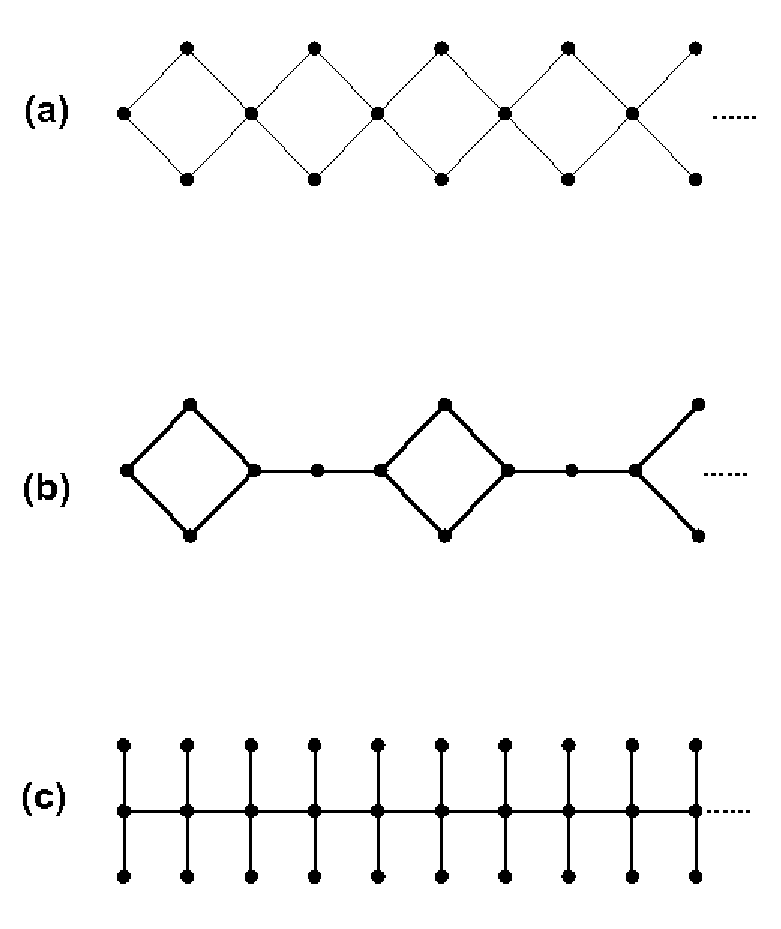}
\vspace{-.4cm} \caption{Examples of bi-regular (panels (a) and (b)) and semi-regular (c) decorated linear chains.} \label{redes} \vspace{-.0cm}
\end{figure}

For illustrative purposes, we first apply this approach to one-dimensional decorated chains in classes $A$ and $B$, which have been extensively used to model polymeric chains \cite{cowie1991, painter1997}. This provides very simple exact results for the dependency between $\psi_{E}(\vec{r})$ and $P_k(E)$, supporting our results for the Apollonian network and percolation clusters. Figure \ref{redes}(a) and (b)  (see, for instance, \cite{macedo1995}) present two examples of networks of class $A$.  Indeed, if in Fig. \ref{redes}(a) the network has $N$ sites, the set $S_1$ comprises $2N/3$ sites with degree $k=2$, while the remaining $N/3$ sites with $k=4$ belong to $S_2$. In Fig. \ref{redes}(b), the lattice with $N$ sites is divided into the set $S_1$, with $3N/5$ sites and degree $k=2$, and the set $S_2$ with $2N/5$ sites and $k=3$. On the other hand, the lattice in Fig. \ref{redes}(c) belongs to class $B$: it is semiregular but non-bipartite, with $N/3$ sites in set $S_1$ (degree $k=4$), and $2N/3$ sites with $k=1$ in set $S_2$.

All eigenvalues and corresponding wave functions (eigenvectors) of the tight-binding Hamiltonian (\ref{eq1}) for each of the simple structures in Fig. (\ref{redes}) can be easily evaluated. Moreover, the periodicity of the structures assures that all wave functions have extended character. The results in the Appendix are valid for the structures in Fig. \ref{redes} (a) and (b). However, the same does not apply to the network in Fig. \ref{redes}(c). For each value of the energy $E(\kappa)$ (here $\kappa$ denotes the one-dimensional quantum wave vector) there exists two different probabilities to finding the particle in a site of the set $S_1$ and $S_2$ respectively. It is straightforward to show that the probability $P_{k} (E)$ is given by

\begin{equation}\label{eq4}
    P_{k=1} (E) = \frac{2}{E^{2}+2},
\end{equation}
\begin{equation}\label{eq5}
    P_{k=4} (E) = \frac{E^{2}}{E^{2}+2}.
\end{equation}

\begin{figure}[h!]
\centering
\includegraphics[width=10cm,angle=0]{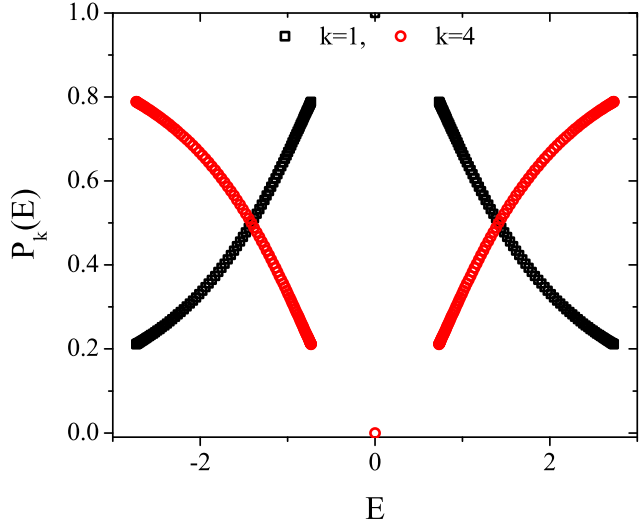}
\vspace{-.8cm}
\caption{Probability of finding the particle on the set of sites with different degree according to Eqs. (\ref{eq4}) ($k=1$, squares) and (\ref{eq5}) ($k=4$ circles) as a function of the wave function energy.} \label{fig2}
\end{figure}

\noindent The above expressions are valid for any of the three possible solutions $E(\kappa)$ of the Schr\"{o}dinger equation derived from the appropriate tight-binding Hamiltonian for the network. Each of the three families of solutions, given by $E(\kappa)=0$, $E=E(\kappa) = cos(\kappa) \pm \sqrt{cos(\kappa)^2 + 2}$, accounts for one third of all possible quantum states. The above equations indicate that the amplitude of the wave function in a certain site depends on its degree. For instance, when $E=0$, all degenerate states satisfy $P_{k=1}(E=0) = 1$ and $P_{k=4}(E=0) = 0$. For the other two energy bands, Fig. \ref{fig2} shows that, when $|E|$ shifts from $\sqrt{3}-1$ towards the band edges at $|E|=\sqrt{3}+1$, $P_{k}(E)$ decreases (increases) for $k=1$  ($k=4$), as a function of $E(\kappa)$. It is astonishing that these qualitative features of this exact result are reproduced in other irregular substrates, including the Apollonian network and the two types of percolation clusters.

To characterize the localization character of the wave function for $E(\kappa) \neq 0$, we evaluate $\xi (E)$ defined by Eq. (\ref{eq3}).  It is straightforward to obtain the expression

\begin{equation}\label{eq6}
    \xi (E)  = \frac{N(E^{2}+2)^{2}}{3(E^{4}+2)}.
\end{equation}

\noindent indicating that $\xi(E) \propto N$ for any energy. Moreover, sites with different values of $k$ have non-zero square modulus of the wave function, which is also delocalized for the ordered structures in Fig. \ref{redes}(c).

The numerical findings discussed in the next sections indicate a similar dependency of $P_k(E)$ with respect to both $k$ and $E$ for more complex systems, we would like to remark that general proofs for the validity of the above observations for any system with more than two types of sites are still required.  Although they are surely beyond the scope of this work, the above discussion sheds light on the interpretation of  some of our results in the next section.

\section{$P_k(E)$ for the Apollonian network}

Now let us discuss the dependence of the probability of site occupation $P_k(E)$ as a function of energy for a geometrical model where the site degree can take integer values in a much wider interval. Despite the fact that several geometrical sets may present this property, we concentrate our investigation on the tight-binding model on the Apollonian network, because most of its quantum energy spectrum features, eigenstate localization properties, and quantum walk dynamics have been described in detail in several works \cite{souza2008,lyra2009,xu2008,souza2013}. For instance \cite{andrade2005}, it is well known that the energy levels are discrete, and also that any eigenvalue that belongs to the energy spectrum having a particular value of $g$ will also be present in the spectra of all $H_{g'}$ with $g'>g$. The energy levels are highly degenerated, and the degeneracy of a level introduced in the spectrum at a given value of $g$ increases with the difference $g'-g$, for $g'>g$.

\begin{figure}[h!]
\centering
\includegraphics[width=10cm,angle=0]{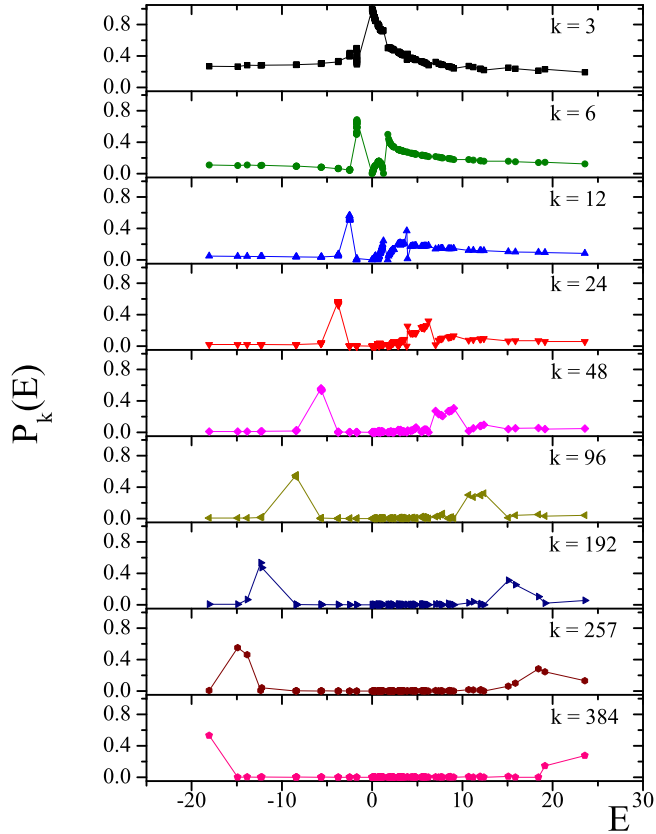}
\vspace{-.8cm}
\caption{Probability of finding the particle on the set of sites with different degree as a function of the wave function energy for the Apollonian network.} \label{fig3}
\end{figure}

Because of the discrete nature of the spectrum, we can not expect to have a smooth dependence of $P_k(E)$ on $E$.  Nevertheless, it is possible to identify that the overall trend displayed in Fig. \ref{fig2} is reproduced in Fig. \ref{fig3}. For the sake of a clarity, we show individual graphs of $P_k(E)$ as a function of $E$ for each possible value of $k$ at $g=8$, when the network comprises 3283 sites.  We see that, for small wave vector $k$, $P_k(E)$ is characterized by large values at small energies $|E|$, and while it becomes small for the levels at large $|E|$. The successive panels indicate that this pattern changes progressively for increasing values of $k$. Large values of $P_k(E)$ can be identified at larger and larger values of $|E|$. If we draw continuous peaks that are upper bounds to the allowed values of $P_k(E)$, we notice that the positions of these peaks are shifted to larger values of $|E|$ as $k$ increases.

We remind that the number of sites with small (large) value of $k$ is large (small). As $P_k(E)$ does not identify particular sites with degree $k$, it is natural that the graphs of $P_k(E)$ are denser (less dense) for small (large) values of $k$. Finally, it is interesting to observe that, as $g'$ increases, the probability of finding a particle at a particular site introduced in the network at generation $g<g'$ is shifted to larger values of $|E|$.

This example shows that is possible to exert control over the location of the electron on sites with different degree $k$ by an adequate choice of the wave-function energy. This is particularly straightforward in the case of the Apollonian network because of the discrete character of its energy spectrum.

\section{The site probability distribution in percolation clusters}

\begin{figure}[h!]
\centering
\includegraphics[width=15cm,angle=0]{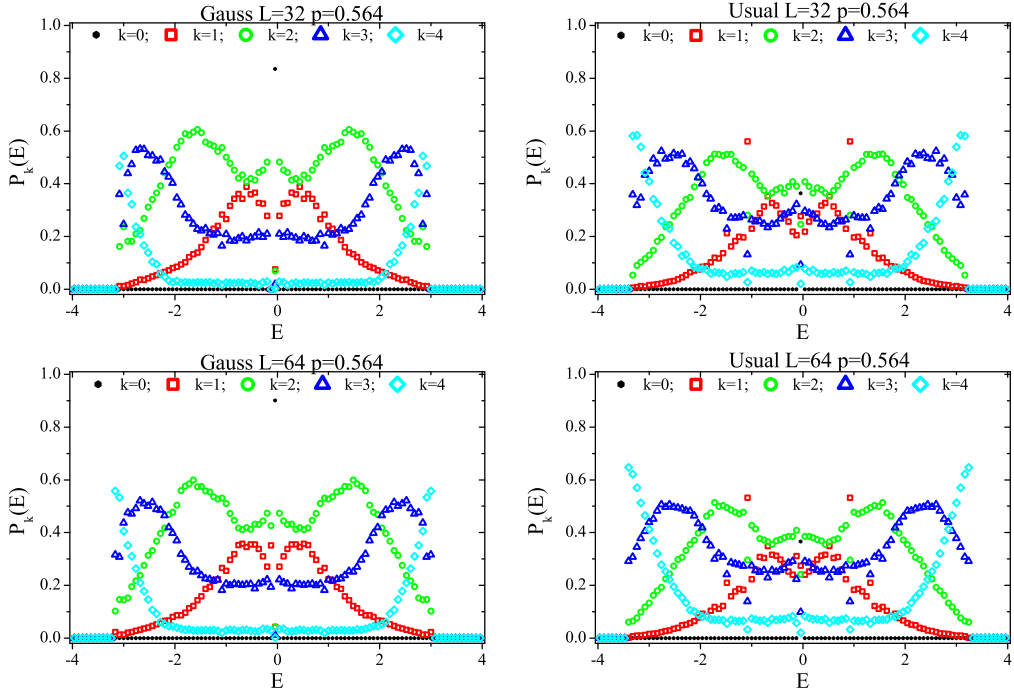}
\vspace{-.8cm}
\caption{Probability of finding the particle on sites with different degree as a function of the wave function energy for percolation clusters obtained by the Gaussian (panels (a) and (c)) and usual models ((b) and (d)) when $p=p_{c,G}=0.56244$. The used samples correspond to $L=32$ (panels (a) and (b)) and $L=64$ ((c) and (d)).}
\label{fig4}
\end{figure}

In this Section we discuss the probability $P_{k} (E)$ of finding the particle on a site of degree $k$, for both the Gaussian and usual random percolation models on the square lattice. Each graph with the results of our numerical simulations consists of five branches for $k=0,1,2,3,4$. Differently from the previous deterministic situations, the results for each model represent averages over $m$ independent samples, obtained according to the same random procedure, and for given values of $L$ and $p$. As we show, depending on the value of these parameters, large fluctuations can still be noticed for $P_{k} (E)$.

\begin{figure}[h!]
\centering
\includegraphics[width=15cm,angle=0]{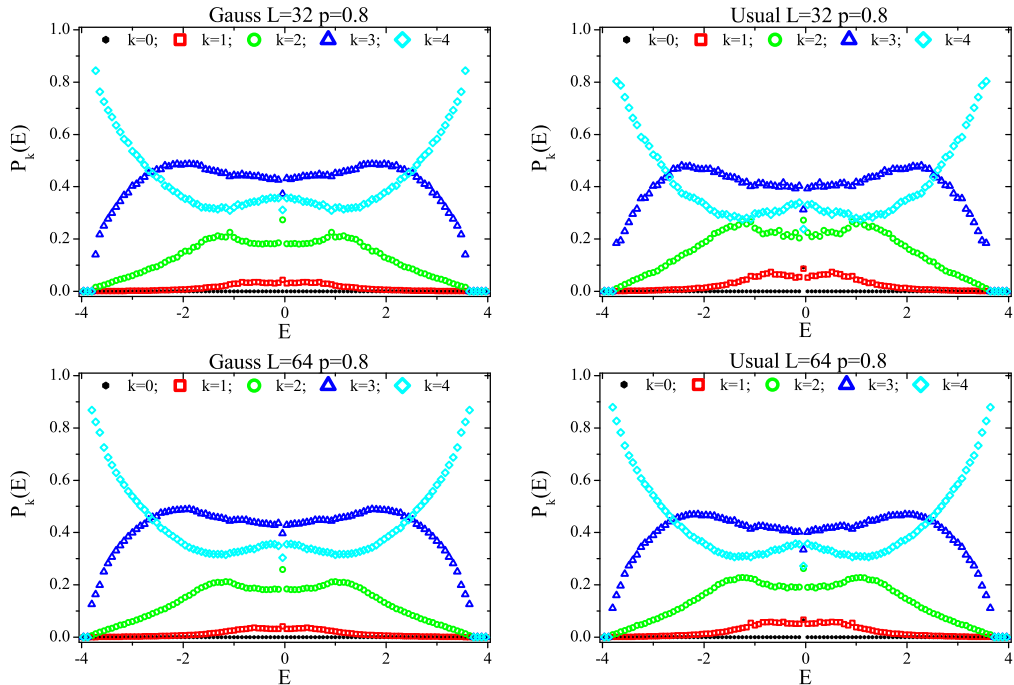}
\vspace{-.8cm}
\caption{Probability of finding the particle on sites with different degree as a function of the wave function energy for percolation clusters obtained by the Gaussian (panels (a) and (c)) and usual models ((b) and (d)) when $p=p_{c,G}=0.8$. The used samples correspond to $L=32$ (panels (a) and (b)) and $L=64$ ((c) and (d)).}
\label{fig5}
\end{figure}

The wave functions have been obtained by numerically evaluating all eigenvalues and eigenvectors of the Hamiltonian (\ref{eq1}). Figures \ref{fig4} and \ref{fig5} show the results for both usual and Gaussian models, for $N=32^2$ and $N=64^2$ and, respectively, bond inclusion probability  $p=0.56244$ and $p=0.8$. The first value of $p$ was chosen to correspond to the percolation threshold  $p_{c,G}$ of the Gaussian model, which is larger than the critical probability $p_{c,U}=0.5$ for usual model. All results were obtained by averaging $P_{k} (E)$ over 10 independently generated samples for each value of $p$.

The comparison of the figures clearly shows that, by decreasing the randomness in the location of bonds, which is achieved by increasing $p$ in the interval $[0.5,1]$, the fluctuations of $P_{k}(E)$ with respect to $E$ become largely damped. Next, we observe that $P_{k}(E)$ associated to each value of $k$ depends on $p$ through the number of sites within each subset. The visual comparison between the average location of points in the two figures indicates that the number of sites in $S_4$ has increased when $p$ changes from $0.564$ to $0.8$, while the opposite is observed for the sets $S_1$ and $S_2$. The situation for $S_3$ is more complex: $P_{k=3}(E)$ increases in the central part of the spectrum, but remains roughly at the same height in the previous peaks situated at $E\simeq\pm 2.2$. The position for the curves for $L=32$ and $L=64$ is almost identical for the Gaussian model, but some differences still can be observed for the usual model. Here, the most important deviation occurs with the branch for $k=2$.

\begin{figure}[h!]
\centering
\includegraphics[width=8cm,angle=0]{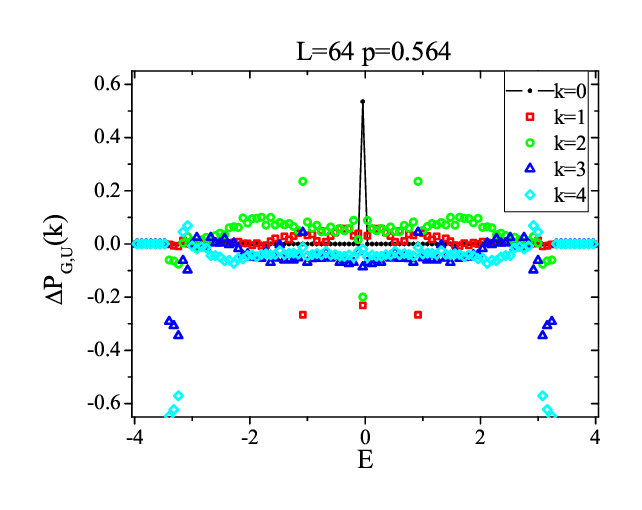}
\includegraphics[width=8cm,angle=0]{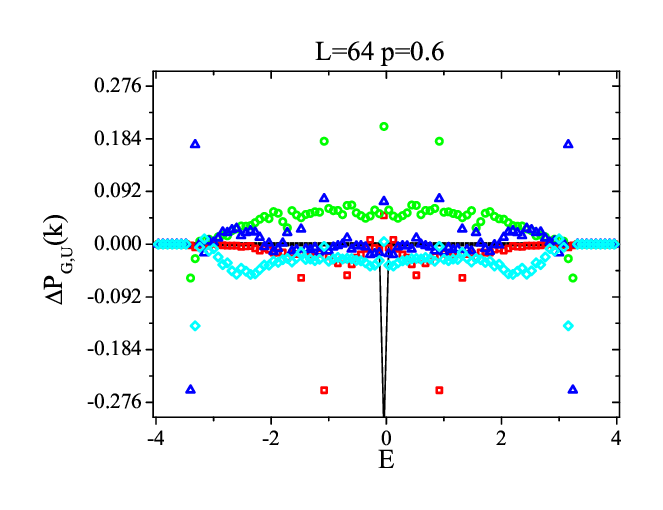}
\includegraphics[width=8cm,angle=0]{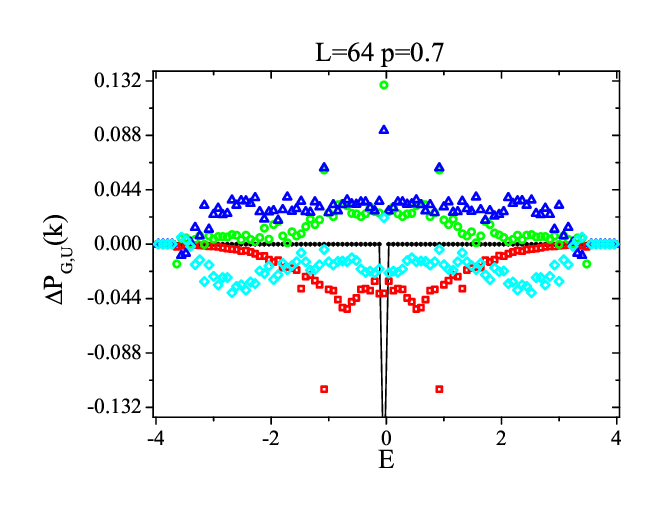}
\includegraphics[width=8cm,angle=0]{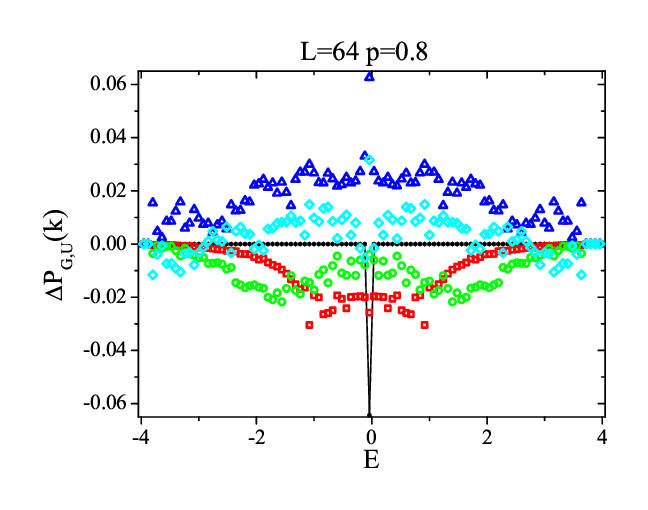}
\includegraphics[width=8cm,angle=0]{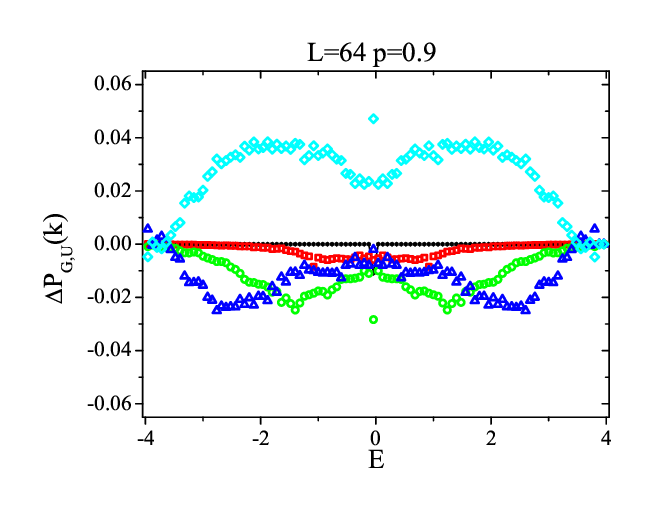}
\vspace{-.8cm}
\caption{Dependence of $\Delta P_{G,U}(k)=P_{k,G}(E)-P_{k,U}(E)$ as a function of $E$ for decreasing probabilities $p=0.56244, 0.6, 0.7, 0.8,$ and $0.9$, when $L=64$. Since $\Delta P_{G,U}(k)$ decreases as $p$ increases, the vertical vertical axis is conveniently re-scaled in each panel. For different $k$, the relative positions of $\Delta P_{G,U}(k)$ also change with $p$.}
\label{fig6}
\vspace{-.0cm}
\end{figure}

The different dependency of $P_{k}(E)$ on $E$ in both models, for $k=3$ and $4$, in Figure \ref{fig5} clearly reminds the behavior shown in Figure \ref{fig2}: the probability for $k=4$ sites is enhanced for energies close to the end of the spectrum, and depleted for the sites with $k=3$. This effect can also be observed for the smaller value $p=0.56244$, as well as for $k=0,1,$ and $2$: all of them decay to zero when $|E|$ increases much faster than the branches for larger values of $k$.

It is possible to identify several distinct features having the form of $P_{k}(E)$ for the two percolation models, which can be better visualized if we consider the differences between the results for the two percolation models for the same values of $L$ and $p$.  The results are illustrated in Figure 6, where we show $\Delta P_{G,U}(k)=P_{k,G}(E)-P_{k,U}(E)$ as a function of $E$, for $L=64$. The additional subscripts $G$ and $U$ identify the Gaussian an usual versions. Substantial differences appear at $p=0.56244$, which gradually decrease when $p$ increases. To make the differences visible, the vertical scales are gradually reduced, so that they differ by a factor ten when we go from $p=0.56244$ to $p=0.8$.  This is an expected dependency, once the percolation clusters generated by the two models attain their largest difference at $p_{c,G}$. Since both models converge to the complete lattice at $p=1$, the differences in the percolation clusters and the localization properties disappear as $p$ increases. However, the reduction in $\Delta P_{G,U}(k)$ is followed by several changes in the position of its $k$-dependent branches as $p$ is increased. For $p=p_{c,G}$, the $k=1$ and $k=2$ branches are positive, while the other are negative. As $p$ increases, the situation is reversed. The $k=1$ and $k=2$ branches become negative, while those for $k=3$ and $k=4$ increase. At $p=0.9$, only the $k=4$ branch is positive. This is a quite interesting behavior of the probability $P_{k}(E)$, revealing that the behavior of the wave function is quite sensitive to differences in the geometric structure of the system.

\section{Conclusions}

We carried out an exhaustive study on the properties of $P_{k}(E)$ as function of $k$ and $E$. Our investigation was based both on analytical results and the numerical evaluation of $P_{k}(E)$ for different types of regular chains, the Apollonian network, and percolation clusters.

We have shown that the probability function introduced/used in this work provides useful information about the energy value of the wave function that should be selected when one aims to enhance the quantum probability on sites with a specific degree. This knowledge may have practical importance, for instance, in the field of quantum information, where we know that information is generated, processed and stored locally on quantum nodes \cite{almeida2013}, or in Josephson  photonic structures, where it is possible to make an analogy between the behavior of quantum particles and wave propagation, in which the frequency response can be controlled by tuning a magnetic field \cite{nori2007}.

One of our analytical results proves that, if the product of $k$ by the number of sites with degree $k$ is the same for any value of $k$ present in the network, the probability of finding a particle in a set of sites with a particular degree $k$ is independent of $k$.

Some specific features following from other exact results for decorated linear chains, regarding the dependence of $P_{k}(E)$ on $k$ and $E$, have also been found in other more complex structures, suggesting that this feature may have a universal character.


\section{Acknowledgements}
We acknowledge financial support from the European Research Council (ERC) Advanced Grant 319968-FlowCCS and the Portuguese Foundation for Science and Technology (FCT) under contracts no. IF/00255/2013, UID/FIS/00618/2013, and EXCL/FIS-NAN/0083/2012.

\section{Appendix}

Here we show that, for a biregular network where the sites can be separated into two distinct sets characterized according to their degree, the product of the site degree of each set by the number of sites in this set is constant.

We start by considering the eigenstates $|\vec{\kappa}\rangle$ of the tight-binding Hamiltonian in a translationally invariant lattice (like in Eq.
(\ref{eq1})) can be expressed by Bloch wave functions according to:
\begin{equation}
    |\vec{\kappa}\rangle =  \frac{1}{\sqrt{N}} \sum_{i=1}^{N} e^{\hat{i}\vec{\kappa}\dot\vec{r}_{i}}
    |\vec{r}_i\rangle.
\end{equation}
\noindent Here $\vec{\kappa}$ is the wave vector of the Bloch function,
$|\vec{r}_i\rangle$ are the local vectors that represent localized functions around the site $\vec{r}_i$, and $N$ indicates the number of sites in the lattice. The eigenstate $|\vec{\kappa}\rangle$ may also be expressed in the vector form
\begin{equation}
    |\vec{\kappa}\rangle =  \frac{1}{\sqrt{N}} (
    e^{\hat{i}\vec{\kappa}\vec{r}_{1}},
    e^{\hat{i}\vec{\kappa}\vec{r}_{2}} , ..., e^{\hat{i}\vec{\kappa}\vec{r}_{N}}
    )^{T}.
\end{equation}

The action of $\mathbf{H}$ on every eigenstate $|\vec{\kappa}\rangle$ is such that
\begin{equation}
    \mathbf{H}|\vec{\kappa}\rangle = (\sum_{\vec{r}'}e^{\hat{i}\vec{\kappa}\vec{r}'} )
    |\vec{\kappa}\rangle = \epsilon_{\vec{\kappa}} |\vec{\kappa}\rangle,
\end{equation}
where $\vec{r}'$ runs over the set of neighbor sites of $\vec{r}_i$, and $\epsilon_{\vec{\kappa}} = \sum_{\vec{r}'}e^{\hat{i}\vec{\kappa}\vec{r}'}$ are the eigenvalues associated with $|\vec{\kappa}\rangle$.

Now let us consider a bi-regular network, like those in Fig. \ref{redes}(a) and \ref{redes}(b). For this case, it is possible to write
\begin{equation}
\mathbf{H}= \sum_{(\vec{r},\vec{r}')} \sum_{\alpha,\alpha'}
|\vec{r}\alpha\rangle \langle \alpha' \vec{r}'|,
\end{equation}
where the system of $N$ sites was divided into $L$ primitive cells labeled by their position $\vec{r}_i$, each one of them containing $n$ sites labeled by the index $\alpha$, in such a way that the wave function, depending both on $\kappa$ and $\alpha$, can be written as
\begin{equation}
    |\psi_{\vec{\kappa}\alpha}\rangle =  \frac{1}{\sqrt{L}} \sum_{i=1}^{L} e^{\hat{i}\vec{\kappa}\vec{r}_{i}}
    |\vec{r}_i \rangle \otimes |\alpha (\vec{\kappa}) \rangle = |\vec{\kappa} \rangle \otimes |\alpha (\vec{\kappa}) \rangle .
\end{equation}

\noindent For this situation,
\begin{equation}
\mathbf{H}= \sum_{\vec{\kappa}} \sum_{\alpha,\alpha'} H_{\vec{\kappa}}(\alpha,\alpha') |\vec{\kappa} \alpha\rangle \langle \alpha' \vec{\kappa}|,
\end{equation}
and
\begin{equation}
    \mathbf{H}|\psi_{\vec{\kappa} \alpha}  \rangle =  \sum_{\alpha'} H_{\vec{\kappa}}(\alpha,\alpha') |\psi_{\vec{\kappa} \alpha'}
     \rangle ,
\end{equation}
where $H_{\vec{\kappa}}(\alpha,\alpha')$ is a $n \times n$ Hermitian matrix which represents the action of $\mathbf{H}$ on the internal states $|\alpha\rangle=(|1\rangle,|2\rangle, ..., |n\rangle)$ of $\vec{\kappa}$-th primitive cell.

The $n$ sites inside the primitive cell of the bi-regular network can be cast into two sets $S_{1}$ and $S_{2}$, containing respectively $n_{S_{1}}$ and $n_{S_{2}}$, in such way that $n_{S_{1}}+n_{S_{2}}=n$. We can choose the basis
$|\alpha\rangle=(|1_{S_1}\rangle, |2_{S_1}\rangle, ...,
|n_{S_1}\rangle,|1_{S_2}\rangle, |2_{S_2}\rangle, ...,
|n_{S_2}\rangle)$ and then $H_{\vec{\kappa}}(\alpha,\alpha')$ can be written in this basis as

\begin{equation*}
H_{\vec{\kappa}}(\alpha,\alpha') = \left(
  \begin{array}{cc}
    \hat{0}_{n_{S_1},n_{S_1}} & \hat{X}^{\vec{\kappa}}_{n_{S_1},n_{S_2}} \\
    \hat{X}^{*\vec{\kappa}}_{n_{S_2},n_{S_1}} & \hat{0}_{n_{S_2},n_{S_2}} \\
  \end{array}
\right),
\end{equation*}

\noindent where $\hat{X}^{\vec{\kappa}}= X(\vec{\kappa}) \hat{1}$, $X(\vec{\kappa})$ is a function that depends on the lattice, and $\hat{0}$ and $\hat{1}$ are the matrices where all elements are respectively set to $0$ and $1$.

We may choose the eigenstates of the primitive cell as
\begin{equation}
|\alpha(\vec{\kappa}) \rangle = \sum_{\alpha =1}^{n_{S_1}}
A_{\vec{\kappa}} |\alpha_{S_1} \rangle + \sum_{\alpha =1}^{n_{S_2}}
B_{\vec{\kappa}} |\alpha_{S_2} \rangle .
\end{equation}

Applying $H_{\vec{\kappa}}(\alpha,\alpha')$ on the $|\alpha(\vec{\kappa}) \rangle$ basis and then diagonalizing, we find the eigen-energies
\begin{equation}
    E_{\vec{\kappa}} = \sqrt{z_{S_1}z_{S_2}}  |X(\vec{\kappa})|,
\end{equation}
\noindent where $z_{S_1}$ and $z_{S_2}$ represent, respectively, the degree of any site in ${S_1}$ and ${S_2}$, and the eigenstate,
by expressing $|\psi_{\vec{\kappa}\alpha}\rangle$ in the vector form, as
\begin{equation}
    |\psi_{\vec{\kappa}\alpha}\rangle =  \frac{1}{\sqrt{L}} (
    e^{\hat{i}\vec{\kappa}\vec{r}_{1}},
    e^{\hat{i}\vec{\kappa}\vec{r}_{2}} , ..., e^{\hat{i}\vec{\kappa}\vec{r}_{L}}
    )^{T} \otimes (A_{\vec{\kappa}}, ..., A_{\vec{\kappa}}, B_{\vec{\kappa}}, ..., B_{\vec{\kappa}})^{T}.
\end{equation}

For $E_{\vec{\kappa}} \neq 0$, we obtain $ |A_{\vec{\kappa}}|^2 =
\frac{1}{2n_{S_1}}$, $ |B_{\vec{\kappa}}|^2 = \frac{1}{2n_{S_2}}$ for all $\vec{\kappa}$. Consequently, the probabilities of finding the electron in the sets $S_1$ and $S_2$ satisfy $p_{S_1}=n_{S_1}|A_{\vec{\kappa}}|^2=1/2=n_{S_2}|B_{\vec{\kappa}}|^2=p_{S_2}$.

\end{document}